\documentclass[12pt]{article}
\usepackage{amssymb}
\usepackage{amsmath, amsthm}
\newcommand{\be}{\begin{equation}}
\newcommand{\ee}{\end{equation}}
\begin{document}
\def\theequation{\arabic{section}.\arabic{equation}}
\begin{titlepage}
\title{Harrison's interpretation of the cosmological redshift 
revisited}
\author{Valerio Faraoni\\\\
{\small \it Physics Department, Bishop's University}\\
{\small \it 2600 College St., Sherbrooke, Qu\'{e}bec, Canada 
J1M~1Z7}\\
{\small \it Email vfaraoni@ubishops.ca}
}
\date{} \maketitle
\thispagestyle{empty}
\vspace*{1truecm}
\begin{abstract}
Harrison's argument against the interpretation of the 
cosmological redshift as a Doppler effect is revisited,  
exaggerated, and discussed. The context, purpose, and 
limitations of the interpretations of this phenomenon are
clarified.  
 \end{abstract}
\end{titlepage} 
\clearpage 
\setcounter{page}{2} 

\section{Introduction}
\setcounter{equation}{0}

There is still much debate  on whether the cosmological redshift 
can be interpreted as a Doppler effect (in the sense of Special 
Relativity) due to the recessional motion of galaxies, as 
originally envisaged by Hubble, or whether this interpretation is 
incorrect \cite{1, 2, 3, 4, 5}. Many current discussions are 
based 
on particular choices of coordinates or observers. We emphasize 
that coordinate-based statements are meaningless in a 
covariant theory such as General Relativity. Families of 
observers, on the contrary, are defined in a 
coordinate-independent way by their four-velocity 
fields (often, 
coordinates and observers are confused because 
geometrically-defined  observers identify certain coordinate 
systems, those in 
which they are at rest, but the distinction should be kept in 
mind). Each observer will have its own  {\em 
interpretation} of  a certain physical phenomenon, which is 
perfectly legitimate {\em for that observer}. A different 
observer will have  a different interpretation, which is 
legitimate as well. It 
is pointless to debate which interpretation is ``the correct 
one'': they are all correct. However, it may happen that strong 
physical reasons select a preferred  family of observers which, 
in turn, selects a preferred interpretation.

The standard textbook  derivation of the cosmological 
redshift for an observer and a light source at rest in comoving 
coordinates in a Friedmann-Lemaitre-Robertson-Walker (FLRW) 
universe does not require the discussion of the 
(four-)velocities  of the source and observer.  However, in 
principle, it does  not prevent one from looking for an 
interpretation in more 
familiar terms, either. The Doppler effect is obtained in 
Special Relativity by a Lorentz transformation from 
a source to an observer in Minkowski space, in which it is 
legitimate to speak of the relative motion and relative velocity 
of a light source and an observer located at different spatial 
points. In a FLRW (or in any curved) space instead one cannot   
compare directly the four-velocities of a source and an observer  
located at different spatial points. It seems, 
therefore, that the interpretation of the cosmological redshift 
in terms of Doppler effect is ill-conceived from the start and 
that this redshift should be  attributed entirely to the 
gravitational field 
of the universe. This interpretation is consistent: it is 
well known that even a  static gravitational field causes 
frequency shifts in null rays propagating through it. The 
classic example of this phenomenon is the shift 
experienced by a photon propagating between a  source and 
an observer both at rest at different radii in the 
Schwarzschild spacetime. 
However, even in this case, one could still give a formal 
representation in 
terms of an ``effective Doppler effect'' by thinking  
of the shift as being equivalent to the Doppler shift that 
would occur if the source (observer) were falling freely to the 
position  of the observer (source), reaching it with non-zero 
velocity.  Narlikar \cite{6} has given a precise meaning 
to this idea; since in a 
curved spacetime one cannot compare 
four-velocities at different points, the best one can 
do is to parallel-transport the four-velocity of the source 
to the observer's  location along the photon worldline,  and then 
construct an {\em effective} Doppler formula at 
that point. This formula contains, as special cases, the Doppler 
effect of Special Relativity, the shift in Schwarzschild space, 
and the cosmological redshift.  It is clear, however, that this 
procedure entails a  fictitious, not a real ``relative motion'', 
especially when this method is applied to the case of 
Schwarzschild space. It seems more convenient to distinguish between 
``purely special-relativistic'' Doppler shift caused by 
local  motion in Minkowski space and ``purely gravitational'' 
shift of  the kind experienced by a photon propagating between a 
source and 
an observer at rest in the Schwarzschild space. It is safe to 
say that, for the latter,  the frequency shift 
is caused by the  fact that the metric tensor $g_{\mu\nu}$ 
assumes different 
values at the different spacetime points where  the source and 
the observer are located. Where does this place cosmological 
redshift then? 
It is easy to see why a naive interpretation of the resdshift in 
terms of Doppler effect tends to linger.  If a source and an 
observer at a fixed comoving distance $d_{comoving}$ are 
locally at rest with respect to the cosmic substratum, they still 
experience what advocates of the Doppler interpretation could 
call a ``relative motion'' in the sense that their physical 
(proper) separation $d_{physical}=a(t) d_{comoving}$ 
(where $a(t)$ is the scale factor) changes with 
time. The derivative of this physical separation with respect to 
the comoving time is 
\be
\dot{d}_{physical}=\dot{a} \, d_{comoving}=H d_{physical} \;,
\ee
{\em i.e.}, Hubble's law, where $H\equiv \dot{a}/a$ is the 
Hubble parameter. It is tempting to interpret $ 
\dot{d}_{physical} $  as a ``velocity 
of recession'' $v$, especially for nearby galaxies, producing a 
redshift factor $v/c$. However, this 
``relative velocity'' is not obtained by comparing directly 
four-velocities at distinct spacetime points: such an operation 
is not defined and giving  a meaning to it requires 
Narlikar's \cite{6} non-local procedure of 
parallel transport. Then, the redshift is unambiguously 
ascribed to the spacetime curvature  and, while Narlikar writes 
down an effective Doppler formula, it is conceptually 
different from the relative motion between source and 
observer in a static Minkowski spacetime. The identification 
of the cosmological reshift factor $z$ with $v/c$, thinking 
of $v$ as a relative velocity in Minkowski space, would be  
arbitrary.

The purpose of this work is to discuss the limits of the 
interpretations of the cosmological redshift in order to gain a 
better understanding  of this
phenomenon.  To this end we develop, 
exaggerate, and discuss an example 
proposed by Harrison \cite{7}, who considered a light 
source and an  observer at rest in Minkowski space.  The source 
emits a light ray  and, while this propagates to the observer, 
space 
suddenly  expands for a short time and then stops expanding 
before 
the observer receives the signal (with the spacetime becoming 
Minkowskian again). Harrison argues that there is cosmological 
redshift given  by the usual formula 
\be \label{redshift}
z+1=\frac{a_O}{a_S}
\ee
 (where $a(t)$ 
is the scale 
factor of the FLRW metric and subscripts~O and~S denote 
observer  and source, respectively) and that, because both 
source and observer are at rest during emission  and detection, 
this  redshift cannot be interpreted as a Doppler effect.
This situation is a special case of the general situation 
contemplated in the derivation of the cosmological redshift. In 
the general situation, consider source and observer widely 
separated in space and time; then, it can be said without risking 
any  interpretation that the redshift is a {\em non-local} effect 
due 
to the different values  of the scale factor at emission and 
absorption. If, following Narlikar \cite{6}, one wants to 
parallel-transport the four-velocity of the source at the 
observer and construct an effective Doppler formula, one is still 
sampling the curvature of  a wide region of spacetime and 
non-local effects. 
This rules out arguments advocating {\em local} motions based 
on the statement that any curved manifold can be locally 
approximated by its tangent space and, therefore, curvature 
effects can only be seen as Doppler effects because curvature is 
not sampled on such small regions. This would be akin to saying 
that the Riemann tensor $R_{\mu\nu\alpha\beta}$  vanishes at one 
point $p$ because space is locally flat while, in fact, 
$R_{\mu\nu\alpha\beta} (p) \neq 0$ but its effects are of higher 
order and are only felt when larger regions of spacetime are 
sampled.  Spacetime is locally flat, but the cosmological 
redshift is a non-local effect due to the curvature of the large 
region of spacetime sampled by the ray between emission and 
detection. A Doppler interpretation based on local arguments is 
meaningless, while one based on Narlikar's parallel transport is 
technically correct, although its convenience is debatable.

We stress that eq.~(\ref{redshift}), together with its 
derivation, is not under discussion here; it 
is a standard and established prediction of the theory, and 
textbook material. What is under 
discussion is its {\em interpretation}.  An interpretation is 
usually part 
of the baggage of a physical theory; the 
interpretation of the cosmological redshift depends on a chosen 
set of observers. {\em A priori}, therefore, different 
interpretations based on different observers are possible and 
they would seem to be equally 
valid.  However, in a spatially homogeneous and isotropic 
universe there  is a physically  preferred set of 
observers--those 
that see the  cosmic microwave background as homogeneous 
and isotropic--they select a preferred interpretation.

Harrison's example has been taken and adapted by 
Saulson to illustrate an analogous situation in a  different 
context \cite{8}. A gravitational wave impinges on a laser 
interferometer and causes a phase shift between the laser beams 
propagating in the two arms. Saulson's purpose is to 
answer a common objection and to demonstrate 
that, although the wavelength of the laser light and the 
interferometer's arm in which it propagates are stretched by 
the same amount, the gravitational wave is observable through 
the non-vanishing phase shift (different approaches to this 
problem can be found in Refs.~\cite{9,10}. Saulson \cite{8} 
correctly notes the 
analogy with Harrison's argument in cosmology and makes his case 
by considering a gravitational wave amplitude described by a 
step function. 

On this line, it  seems fit to revisit Harrison's argument 
and exaggerate it for clarity by assuming that the 
cosmic expansion  takes place suddenly at a single instant of 
time. For  simplicity,  the spacetime metric is given by the 
spatially flat FLRW line element 
\be\label{metric}
ds^2=-dt^2+a^2 (t)\left( dx^2 +dy^2 +dz^2 \right)
\ee
in comoving coordinates $\left( t, x,y,z \right)$ and with scale 
factor $a(t)=1+\theta(t)$, where 
\be 
\theta(t)= \left\{ 
\begin{array}{c}
0 \;\;\;\;\;\;\;{\mbox if} \;\;\;\; t<0 \\
  \\
1 \;\;\;\;\;\;\;{\mbox if} \;\;\;\; t \geq0 \\
\end{array} \right.
\ee
is the Heaviside step function. This describes a universe 
doubling its size abruptly at time $t=0$.

There are risks in adopting discontinuous metrics: the 
gravitational accelerations (Christoffel symbols) are 
impulsive, while the 
curvature tensor and its contractions the Ricci tensor and 
Ricci scalar may be ill-defined because, in general, they 
contain products of distributions. At best, the corresponding 
stress-energy tensors will also be distributional. In spite of 
these formal difficulties, discontinuous and even delta-like 
metrics have been considered long ago by Penrose \cite{11}.  
Such exact solutions of the Einstein  equations can be obtained 
by cut-and-paste procedures in which different regions of 
spacetime (perhaps Minkowski spaces, as in our present example) 
are joined together with a suitable warp along an 
hypersurface \cite{11}. 
Similar metrics have been studied extensively in the  literature 
on  exact plane gravitational waves \cite{12} and it has also 
been discussed how to make sense of the product of distributions 
in  General Relativity \cite{13}.  It is not too surprising, 
therefore, 
to  see a discontinuous metric in Saulson's example 
which deals with gravitational waves \cite{8}. 
Abstracting from the technical difficulties with the Riemann 
tensor,  in the 
following we  study the propagation of a null ray between a 
source and an 
observer which are at rest initially (when light is emitted) and 
after the expansion of the universe has stopped (when light is 
received by 
the observer)--we derive and discuss the corresponding 
redshift formula.   The calculation does not require 
the  consideration of the curvature.

\section{Harrison's model revisited}
\setcounter{equation}{0}

The metric tensor and its inverse are
\begin{equation}
\left( g_{\mu\nu} \right)=\left(
\begin{array}{cccc}
-1 & 0 & 0 & 0 \\
& & &\\
0 & 1+3\theta & 0 & 0 \\
& & &\\
0 & 0 & 1+3\theta & 0 \\
& & &\\
0 & 0 & 0 & 1+3\theta 
\end{array} \right) \;, \;\;\;\;\;\;\;\;
\left( g^{\mu\nu} \right)=\left(
\begin{array}{cccc}
-1 & 0 & 0 & 0 \\
& & &\\
0 & \frac{1}{1+3\theta} & 0 & 0 \\
& & &\\
0 & 0 & \frac{1}{1+3\theta} & 0 \\
& & &\\
0 & 0 & 0 & \frac{1}{1+3\theta} 
\end{array} \right) \;, 
\ee
respectively. The  only 
non-vanishing Christoffels symbols are 
\begin{eqnarray}
&& \Gamma^0_{11}=\Gamma^0_{22}=\Gamma^0_{33}=\frac{3 \, 
\delta(t)}{2}  \;, \nonumber \\
&& \nonumber \\
&& 
\Gamma^1_{01}=\Gamma^1_{10}=\Gamma^2_{02}=\Gamma^2_{20}=
\Gamma^3_{03}=\Gamma^3_{30}=\frac{3 \, \delta(t)}{2\left[ 
1+3\theta(t) \right]} \;,
\end{eqnarray}
where $\delta (t)$ denotes the Dirac delta. 

Consider now a light source located at $x=L$ and an observer at 
$x=0$, both on the $x$-axis. The source emits a signal at time 
$t_S$, which  is received at time $t_O$ by the observer, with 
$t_S <0< t_O$. The null ray propagates along the  
$x$-axis in the direction of decreasing $x$ and has four-tangent
\be
u^{\mu}=\frac{dx^{\mu}}{d\lambda}=\left( u^0, u^1, 0, 0 \right) 
\;, 
\ee
where $\lambda $ is a parameter along the null geodesic. The 
normalization $g_{\mu\nu}u^{\mu}u^{\nu}=0$ yields
\be \label{normaliz}
u^1=-\, \frac{u^0}{\sqrt{ 1+3\theta(t)}} \;.
\ee
The negative sign is chosen because $u^0=dt/d\lambda >0$ 
corresponding to the parameter $\lambda$  increasing along the 
null geodesic, and the ray propagates in the direction of 
decreasing $x$. The zero component of the null geodesic 
equation
\be
\frac{ du^{\mu} }{ d\lambda } 
+\Gamma^{\mu}_{\alpha\beta} u^{\alpha} u^{\beta} =0 
\ee
then yields
\be
\frac{du^0}{d\lambda}=- \Gamma^{0}_{11}\left( u^0 \right)^2 =
-\, \frac{3\delta(t) ( u^0 )^2}{2\left[ 1+3\theta(t) \right] } 
\;.
\ee
By using the definition of $u^0$, it is seen that  
$d\lambda=dt/u^0$ and 
\be
\frac{du^0}{dt}=- \, \frac{3 u^0  \delta(t)}{2\left[ 
1+3\theta(t) \right] } 
\;.
\ee
The integration of this equation along the null geodesic between 
the 
source~S and the observer~O yields\footnote{Note that one cannot 
simply use the property 
$\int_{t_S}^{t_O} f(t) \delta (t) dt= f(0)$ because $f(t)=\left[ 
1+3\theta (t) \right]^{-1}$ is not  a test function continuous 
with all its derivatives; the integration is nevertheless 
straightforward.} 
\be
\ln \left[ \frac{u^0_{(O)}}{ u^0_{(S)} } \right] =-\frac{3}{2} 
\int_{t_S}^{t_O}dt\, \frac{\delta(t)}{1+3\theta(t)}=-\ln 2 \;,
\ee
therefore,
\be
u^0_{(O)}= \frac{ u^0_{(S)} }{2} \;.
\ee
Light is emitted in flat spacetime and the four-tangent to the 
null ray at~S is $u^{\mu}_{(S)}=\left( 1, -1, 0, 0 \right)$, 
hence
\be
u^0_{(O)}=\frac{1}{2} \;.
\ee
The propagating photon has four-wavevector $k^{\mu} =\omega 
u^{\mu} =\left( \omega, \vec{k} \right)$ in flat spacetime. For 
a photon emitted at the source with unit angular frequency, it 
is $k^{\mu}=u^{\mu}$ and the angular frequency measured by the 
observer is $k^0_{(O)}=u^0_{(O)}=1/2$. In fact, the angular 
frequency measured by any observer with four-velocity 
$v^{\mu}$ is $-k^{\mu}v_{\mu}$, and both source and 
observer have four-velocity $v^{\mu}=\delta^{0\mu}$ in comoving 
coordinates.  A photon emitted at~S 
with angular frequency $\omega_S$ will have an angular frequency 
as measured by~O
\be
\omega_O=\frac{\omega_S}{2} \;,
\ee
{\em i.e.}, light is  redshifted by the sudden cosmic 
expansion  at $t=0$. The redshift factor defined in 
terms of the wavelenghts $\lambda^{e.m.}_{S,O}$ of the 
electromagnetic signal at the source and 
observer is 
\be
z\equiv \frac{\lambda^{em}_O-\lambda^{em}_S}{\lambda^{em}_S}=\frac{
\lambda^{em}_O}{\lambda^{em}_S}-1=\frac{\omega_S}{\omega_O}-1=1 \;.
\ee
Note that $z+1=2$ and that the universe has doubled its size at 
$t=0$, so the usual formula for the cosmological 
redshift~(\ref{redshift})  is satisfied. As a check of this 
little calculation, one can integrate also the 
$x$-component of the null geodesic equation 
\be
\frac{du^1}{d\lambda}=-2\Gamma^1_{01}u^0 u^1 = \frac{ 
3\delta(t) (u^1)^2}{\sqrt{ 1+3\theta(t)}} \;.
\ee
The integration between~S and~O as above yields
\be
u^1_{(O)}=\frac{ u^1_{(S)}}{4} 
\ee
and, using $u^{\mu}_{(S)}=\left( 1, -1, 0, 0 \right)$, it is 
$u^1_{(O)}=-1/4$. This value of $u^1_{(O)}$ coincides with the 
one  obtained from eq.~(\ref{normaliz}) evaluated at~O, {\em 
i.e.}, $ u^1_{(O)}=-u^0_{(O)}/2=-1/4$.

\section{Discussion}
\setcounter{equation}{0}

The cosmological redshift factor~(\ref{redshift}) is 
a straightforward prediction of standard cosmology, but its {\em 
interpretation} seems to be still controversial, judging from 
the number of articles debating it. Our implementation of 
Harrison's example should shed some light here. 
According to Harrison \cite{7} the cosmological redshift 
cannot be interpreted as a Doppler shift because both source and 
observer are at rest when the signal is emitted or received. 
This redshift is due to the fact that the scale factor (the 
only degree of freedom of the metric $g_{\mu\nu}$) assumes 
different values at the spacetime 
points~S and~O. Narlikar's procedure of 
parallel-transporting the source four-velocity along the null 
geodesic from~S to~O and constructing an effective 
non-local Doppler 
formula there \cite{6}  can be applied, but it is not 
particularly 
useful for understanding the physics involved: here we do 
have a source and an observer at rest in comoving 
coordinates. One could  object that, after all,  these source and  
observer 
are not truly at rest all the time: the comoving 
distance between S and~O is $L$, while their physical (proper)  
separation 
\be
d_{physical}(t)=a(t) L= \left\{ \begin{array}{c}
L \;\;\;\;\; {\mbox if } \;\;\;\; t<0\\
 \\
2L \;\;\;\;\; {\mbox if } \;\;\;\; t \geq 0
\end{array} \right.
\ee
doubles suddenly at $t=0$ with infinite ``velocity'' $ \frac{d 
d_p}{dt}=L \delta(t)$. Here 
we {\em assumed} that source and observer return to rest after 
the universe has expanded, without worrying about whether this 
happens spontaneously or some entity forces them to do so. It is 
shown in the Appendix that they actually remain spontaneously 
at rest.

In some sense, therefore, there has been some ``relative motion'' 
of~S and O, but this is not a local motion in the sense of 
Special Relativity. The space in which~S and~O live has 
expanded, changing their proper separation, and this sudden 
``non-local motion'' has redshifted the light.

It has been  argued \cite{1, 4}  that the 
interpretation  depends on  the 
coordinate or gauge adopted; it is more correct to say that it 
depends on the {\em set of observers} adopted (the latter are 
defined in a coordinate-independent way by their 
four-velocities).   It is certainly 
true that different observers, which define different coordinate 
systems (those in which they are at rest with their 
four-velocity components given by $v^{\mu}=\delta^{0\mu}$), will 
have 
different  interpretations of the same physics. Each one of this 
is a legitimate interpretation {\em for that observer.}  
While the redshift factor is an observer-invariant quantity, 
its {\em interpretation} is not. There are other examples in 
which, by 
choosing different observers, coordinate systems, or gauges, 
different interpretations of the same physics arise. In 
a laser-interferometric detector of gravitational waves one 
can choose a gauge in which the end mirrors move, and attribute 
the phase shift to the relative motions of the mirrors, which 
differ  in the two arms; or one can choose the 
transverse-traceless (TT) gauge in which the mirrors are at rest 
in TT coordinates (but their  proper separation changes) and 
attribute the phase shift to the  different rates 
at which time elapses (a frequency shift effect again) and 
caused by the 
passing gravitational wave. Of course, the physics is 
gauge-invariant: the phase shift (a scalar, therefore  
gauge-invariant,  
quantity) is ultimately a tidal effect caused by the curvature 
tensor, which is also a gauge-invariant quantity. Another 
example is quantum mechanics, in which the Schr\"odinger, 
Heisenberg, and interaction pictures all provide different 
interpretations of the same physics, and the use of these  
different formalisms is a matter of convenience.

The cosmological redshift in  a FLRW space is a geometric, 
non-local,  
coordinate-invariant effect. It seems that there is little point 
arguing in favour of one of  its interpretations based on 
a set of observers versus 
another. However, in a FLRW space, there  definitely is a set of 
physically preferred  observers: they are the comoving observers 
who see the cosmic  microwave background homogeneous 
and isotropic around them (apart from small temperature 
fluctuations). It is arguable  whether an interpretation of 
the redshift  formula~(\ref{redshift}) derived in such a clear 
way  in  textbooks is really necessary but, if one opts for 
choosing 
an interpretation, this should be tied to the physically 
preferred comoving observers. For the latter,  the 
redshift is  definitely gravitational and  not due to an 
ill-defined recessional motion of galaxies.

There is still a difficulty to be dealt with. In Ref.~\cite{1} 
discussing the interpretation of the cosmological redshift, the 
Milne universe is considered as an example: this 
is a region of Minkowski space written in accelerated 
coordinates, which yields the line element of an open universe 
with linear scale factor $a(t)=t$,
\be
ds^2=-dt^2+t^2 \left[ d\chi^2 +\sinh^2 \chi \left( d\theta^2 
+d\varphi^2 \right) \right] \;.
\ee
Of course, a calculation of the Riemann tensor shows that it 
vanishes everywhere and that this is (a portion of) flat 
spacetime foliated  using hyperbolic 3-surfaces.\footnote{The 
spatial curvature is negative, but the spacetime 
curvature is zero.}  There  is 
redshift, which is certainly not due to gravity (which is absent 
here); it is definitely a Doppler shift due to the fact  
that the observers at rest in $\left( t, \chi, \theta, \varphi 
\right)$  coordinates (``Milne observers'') are moving away from 
each other when seen from observers at rest in Minkowski space. 
The latter are comoving observers, but perhaps one could say 
that also the Milne observers can be called ``comoving''; this 
ambiguity in the term ``comoving'' arises because spacetime is 
empty and {\em any} observer can be said to be ``comoving'' with 
no matter. However, this is really a pathological example and 
its choice to discuss the interpretation of the cosmological 
redshift is an unhappy one. A set of geometrically preferred 
observers consists of those with four-velocity $v^{\nu}$ 
parallel to the timelike Killing field of this metric, 
which is defined in an invariant way.

If the energy density $\rho$ of a 
FLRW universe is non-zero, there will be precisely one set of 
physically preferred observers that see zero spatial gradients 
of $\rho$ (comoving  observers) and the ambiguity in the 
interpretation of the  cosmological redshift (gravitational 
versus Doppler) disappears.

\section*{Acknowledgments}

This work is supported by the Natural Sciences and Engineering
Research Council of Canada. 

\clearpage
\section*{Appendix}

Here we address the question of whether two objects initially at 
rest remain at rest or are set in relative motion after the 
expansion of the universe has stopped. Consider an observer 
with timelike four-velocity $v^{\mu}$ satisfying the initial 
condition $v^{\mu}_{initial} =\left( 1, 0,0,0 \right)$ at 
$t<0$ and the normalization  $g_{\mu\nu} v^{\mu} 
v^{\nu}=-1$  (this could be the source~S or the observer~O). The 
geodesic equation yields
\begin{eqnarray}
&& \frac{dv^0}{d\lambda}=-\left[ \Gamma^0_{11} ( v^1)^2 +
\Gamma^0_{22} ( v^2)^2 + \Gamma^0_{33} ( v^3)^2 \right]= -\, 
\frac{ 3 (\vec{v} )^2 \delta(t)}{2} \;, \\
&& \nonumber \\
&& \frac{dv^i}{d\lambda}= -\, 
\frac{ 3 \delta(t) v^0 v^i}{1+3\theta(t) } \;,
\end{eqnarray} 
where $(\vec{v})^2=(v^1)^2 +(v^2)^2 + (v^3)^2 $.  The 
normalization of $v^{\mu}$ yields
\be\label{questa}
(\vec{v})^2=\frac{ (v^0)^2-1}{1+3\theta(t)} \;.
\ee
Using eq.~(\ref{questa}) and $d\lambda=dt/v^0$ along the 
timelike geodesic with tangent $v^{\mu}$, and integrating
 between times  $t_{initial}$ and $t_{final}$, one obtains
\be
v^i_{final}=\frac{ v^i_{initial}}{1+3\theta(t)}=\left\{
\begin{array}{c}
v^i_{initial} \;\;\;\;\;\;\; {\mbox if} \;\;\; t<0 \\
\\
\frac{ v^i_{initial}}{4}  \;\;\;\;\;\;\; {\mbox if} \;\;\; t 
\geq 0 
\end{array} \right.
\ee
The initial condition $v^i_{initial}=0$ then guarantees that 
$v^i=0$ at all times.  Indeed, due to spatial isotropy, an object 
initially at rest cannot pick up a spatial velocity as a 
consequence of  the sudden 
cosmic expansion because this would select a preferred direction 
in 3-space.


\end{document}